\documentclass[prd,floatfix,onecolumn,amsmath,amssymb,floatfix]{revtex4}
\usepackage{graphicx,color,dcolumn,booktabs,bm}
\usepackage{subfigure}
\usepackage{longtable,lscape}
\usepackage{amssymb}
\usepackage{indentfirst}
\usepackage{epsfig}
\usepackage{feynmf}   
\usepackage{epstopdf}   
\usepackage{slashed}  
\usepackage{cases}
\usepackage[pdfpages]{xcolor}
\definecolor{maroon}{RGB}{139,25,150}
\usepackage{multirow}
\usepackage{float}
\usepackage{graphicx,color,dcolumn,booktabs,bm}
\usepackage[colorlinks, citecolor=blue,anchorcolor=red,menucolor=red, linkcolor=red,filecolor=red,runcolor=red,urlcolor=blue,frenchlinks=red, urlcolor=blue]{hyperref}

\begin{document}

\preprint{}
\preprint{}
\title{\color{maroon}{Interpretation of  the $\Lambda_c(2910)^+$ baryon newly seen by  Belle Collaboration and its possible bottom partner}}

\author{K.~Azizi}
\affiliation{Department of Physics, University of Tehran, North Karegar Avenue, Tehran
14395-547, Iran}
\affiliation{Department of Physics, Do\u gu\c s University,
Ac{\i}badem-Kad{\i}k\"oy, 34722 Istanbul, Turkey}
\author{Y.~Sarac}
\affiliation{Electrical and Electronics Engineering Department,
Atilim University, 06836 Ankara, Turkey}
\author{H.~Sundu}
\affiliation{Department of Physics, Kocaeli University, 41380 Izmit, Turkey}

\date{\today}

\begin{abstract}
The developments in the experimental facilities and analyses techniques have recently  lead to the observation of many hadronic states ranging from excitations of conventional hadrons  to various exotic states. The baryons with single heavy quark are among these states providing an attractive field of research to get a better understanding of the nonperturbative nature of the strong interaction. Recently, the Belle Collaboration announced observation of  the  state $\Lambda_c(2910)^+$ with a mass  $2913.8\pm5.6\pm3.8~\mathrm{MeV}/c^2$ and width  $51.8\pm20.0\pm18.8~\mathrm{MeV}$.  In the present study,  by the mass analyses  of different excitations at $\Lambda_c$ channel and their comparison with existing experimental information, we find that  the  spin-parity of this newly found excited state is  $ J^P= \frac{1}{2}^-$ and it is a $ 2P $ state denoting by  $\Lambda_c(\frac{1}{2}^-,2P)$. We predict its current coupling as well, which can be served as one of the main input parameters to investigate different decays and interactions of this particle. We also determine the mass and current coupling  of  $\Lambda_b(\frac{1}{2}^-,2P)$ as possible  bottom counterpart  of the new  $\Lambda_c(2910)^+$ state, which may be in agenda of different experiments in near future. 

\end{abstract}

\maketitle
\section{Introduction}

With the developed experimental facilities and techniques, we have recently witnessed the observation of various hadronic states ranging from excited states of conventional baryons with heavy quark content to some exotic states. These observations have boosted theoretical studies over these states either to provide an explanation for their observed properties, such as their spectroscopic parameters or possible quantum numbers, or to offer other possibly present such states for experimental investigations. Among these theoretical researches are the spectroscopic analyses of the baryons containing a single heavy quark, which provide an excellent opportunity to investigate the dynamics of light quarks in the presence of the heavy ones and to improve the understanding of the nonperturbative nature of the quantum chromodynamics (QCD). Delving into these states also provides a test for the predictions of the quark model, the heavy quark symmetry, and the other theoretical models working over the properties of these states.

The last few decades have become an era in which various ground or excited states of the baryons containing heavy quarks have joined the baryon family via experimental observations. Although plenty of these baryons were observed and listed in the Particle Data Group (PDG)~\cite{ParticleDataGroup:2020ssz}, there are still missing ones awaiting to be observed, and some of the observed ones are in need of a better understanding of their poorly known properties. Since the first observation of the ground state $\Lambda_c^+$ baryon by Fermilab in 1976~\cite{Knapp:1976qw}, almost all the ground states of singly heavy baryons were observed, and many of their excited states were detected. These states with single heavy quark content include $\Lambda_c(2595)$~\cite{ARGUS:1993vtm,CLEO:1994oxm,ARGUS:1997snv}, $\Lambda_c(2625)$~\cite{E687:1993bax,CLEO:1994oxm}, $\Lambda_c(2765)^+$ and $\Lambda_c(2880)^+$~\cite{CLEO:2000mbh},  $\Lambda_c(2940)^+$~\cite{BaBar:2006itc}, $\Lambda_c(2860)^+$~\cite{LHCb:2017jym}, $\Sigma_c(2520)$~\cite{Ammosov:1993pi,CLEO:1996czm}, $\Sigma_c(2800)$~\cite{Belle:2004zjl}, $\Xi_c(2645)$~\cite{CLEO:1995amh,CLEO:1996zcj,E687:1998dwp}, $\Xi_c(2790)$~\cite{CLEO:2000ibb}, $\Xi_c(2815)$~\cite{CLEO:1999msf}, $\Xi_c(2923)$~\cite{LHCb:2020iby}, $\Xi_c(2930)$~\cite{BaBar:2007xtc}, $\Xi_c(2970)$~\cite{Belle:2006edu}, $\Xi_c(3055)$~\cite{BaBar:2007zjt}, $\Xi_c(3080)$~\cite{Belle:2006edu}, $\Xi_c(3123)$~\cite{BaBar:2007zjt}, $\Omega_c(2770)^0$~\cite{BaBar:2006pve}, $\Omega_c(3000)^0$, $\Omega_c(3050)^0$, $\Omega_c(3065)^0$, $\Omega_c(3090)^0$ and $\Omega_c(3120)^0$~\cite{LHCb:2017uwr} with single charm quark. And the states containing single heavy bottom quark  are $\Lambda_b(5912)^0$, $\Lambda_b(5920)^0$~\cite{LHCb:2012kxf}, $\Lambda_b(6070)^0$~\cite{LHCb:2020lzx}, $\Lambda_b(6146)^0$, $\Lambda_b(6152)^0$~\cite{LHCb:2019soc}, $\Sigma_b(6097)^+$, $\Sigma_b(6097)^-$~\cite{LHCb:2018haf}, $\Xi_b(6100)^-$~\cite{CMS:2021rvl},  $\Xi_b(6227)^-$~\cite{LHCb:2018vuc},  $\Xi_b(6227)^0$~\cite{LHCb:2020xpu}, $\Omega_b(6316)^-$, $\Omega_b(6330)^-$, $\Omega_b(6340)^-$ and $\Omega_b(6350)^-$~\cite{LHCb:2020tqd}. Some of these states need further experimental investigation since the evidence for their existence is poor, and some others, such as $\Lambda_c(2765)^+$ or $\Sigma_c(2765)$, $\Sigma_c(2800)$, $\Xi_c(2923)$, $\Xi_c(2930)$,$\Xi_c(3055)$, $\Xi_c(3080)$, $\Xi_c(3123)$, $\Omega_c(3000)^0$, $\Omega_c(3050)^0$, $\Omega_c(3065)^0$, $\Omega_c(3090)^0$, $\Omega_c(3120)^0$ and $\Lambda_b(6070)^0$, $\Sigma_b(6097)^+$, $\Xi_b(6227)^-$, $\Xi_b(6227)^0$, $\Omega_b(6316)^-$, $\Omega_b(6330)^-$, $\Omega_b(6340)^-$ and $\Omega_b(6350)^-$~\cite{ParticleDataGroup:2020ssz},  have not well determined quantum numbers. Therefore their investigation is an active field both from experimental and theoretical respects.

Based on either the expectation of their presence or the reports of their experimental observation, these baryons were widely investigated, and their masses or decay mechanisms were scrutinized in the framework of various theoretical models. These models include quark model~\cite{Copley:1979wj,Ivanov:1998wj,Ivanov:1999bk,Hussain:1999sp,Maltman:1980er,Capstick:1985xss,Capstick:1986bm,Ebert:2005xj,Valcarce:2008dr,Ebert:2011kk,Karliner:2015ema,Yoshida:2015tia,Shah:2016mig,Shah:2016nxi,Thakkar:2016dna,Albertus:2005zy,Migura:2006ep,Zhong:2007gp,Ebert:2007nw,Garcilazo:2007eh,Roberts:2007ni,Hernandez:2011tx,Liu:2012sj,Chen:2016iyi,Nagahiro:2016nsx,Wang:2017kfr,Chen:2018vuc,Wang:2018fjm,Yao:2018jmc,Wang:2019uaj}, relativistic flux tube model~\cite{Chen:2014nyo}, $^3P_0$ model~\cite{Chen:2007xf,Ye:2017dra,Chen:2017aqm,Ye:2017yvl,Yang:2018lzg,Guo:2019ytq,Lu:2019rtg,Liang:2019aag}, heavy hadron chiral perturbation theory~\cite{Huang:1995ke,Banuls:1999br,Cheng:2006dk,Cheng:2015naa,Jiang:2015xqa}, lattice QCD~\cite{Padmanath:2013bla,Bali:2015lka,Bahtiyar:2015sga,Bahtiyar:2016dom}, the bound state picture~\cite{Chow:1995nw}, Bethe-Salpeter formalism~\cite{Guo:2007qu}, QCD sum rules~\cite{Zhu:2000py,Wang:2010it,Mao:2015gya,Chen:2016phw,Mao:2017wbz,Wang:2017vtv,Agaev:2017jyt,Agaev:2017lip,Aliev:2018lcs,Aliev:2018ube,Cui:2019dzj,Azizi:2020tgh,Azizi:2020ljx,Azizi:2020azq,Agaev:2020fut}, and the light cone QCD sum rules~\cite{Zhu:1998ih,Wang:2009ic,Wang:2009cd,Aliev:2009jt,Aliev:2010yx,Aliev:2014bma,Aliev:2016xvq,Chen:2017sci,Agaev:2017ywp,Aliev:2018vye}.  One can also refer to the Refs.~\cite{Richard:1992uk,Korner:1994nh,Klempt:2009pi,Crede:2013sze,Cheng:2015iom,Chen:2016spr} and the references therein for more discussion of the properties of these states.   

As is stated, deeper understanding of the properties such as their spectroscopic parameters or decay mechanisms to get information about the possible quantum numbers of the observed baryons with a single quark improves our understanding of the strong interaction at low energy regions. With this motivation, in Refs.~\cite{Aliev:2018lcs,Aliev:2018vye,Azizi:2020tgh,Azizi:2020ljx,Azizi:2020azq} we investigated some of the excited states of the single heavy baryons, namely $\Xi_b(6227)^-$, $\Sigma_b(6097)^{\pm}$, $\Lambda_b(6146)^0$, $\Lambda_b(6072)^0$, and $\Xi_b(6227)^0$, with the aim of fixing their quantum numbers. To this end, we applied either the analyses of spectroscopic parameters or their decay channels and made the analyses using one of the effective nonperturbative approaches, the QCD sum rules method~\cite{Shifman:1978bx,Shifman:1978by,Ioffe81}. The method applied has proven its success many times with its reliable predictions that are consistent with experimental observations up to date. In this work, inspired by the recent observation of the Belle Collaboration reporting a new structure $\Lambda_c(2910)^+$ in the invariant mass spectra, $M_{\Sigma_c(2455)^{0,++}\pi^{\pm}}$, we investigate the $2P$ excited states of the $\Lambda_c^+$ and  $\Lambda_b^0$ baryons and compare our findings for these excited states with this observation. For the observed new state, the measured mass was presented as $(2913.8\pm5.6\pm3.8)~\mathrm{MeV}/c^2$, and its width was reported as $(51.8\pm20.0\pm18.8)~\mathrm{MeV}$~\cite{Belle:2022hnm}. Moreover, the observed state has been suggested to be a possible candidate for $\Lambda_c(\frac{1}{2}^-,2P)$. After the observation of the $\Lambda_b(6072)^0$ by the LHCb Collaboration~\cite{LHCb:2020lzx}, offering the observed state as a  $2S$ excitation of the $\Lambda_b^0$ baryon, we investigated spin-$\frac{1}{2}$ $\Lambda_b^0$ and $\Lambda_c^+$ baryons in Ref.~\cite{Azizi:2020ljx}. Our predictions in the mentioned work for the masses of $1S$, $1P$, and $2S$ states which are also presented in the Sec.~\ref{Sec3} of this work were lower than that of the present observation of the Belle Collaboration. This indicates that the interpretation for $\Lambda_c(2910)^+$ state as one of the spin-$\frac{1}{2}$ $1P$ or $2S$ excitations of $\Lambda_c^+$ state may not be suitable. Therefore, in this work, we aim to investigate the possible quantum numbers of $\Lambda_c(2910)^+$ state in light of the suggested quantum numbers in Ref.~\cite{Belle:2022hnm} and its possible bottom counterpart. To achieve this aim, we use the QCD sum rule method for the calculations of the mass and current coupling constant of the considered state. In the QCD sum rule calculation, a proper interpolating current is chosen with the same quantum numbers and quark content as the considered state. This interpolating current can also create or annihilate any state having the same quantum numbers and quark contents as the considered state, including its excitations. Therefore the ones chosen in this work couple with the low-lying states of spin-$\frac{1}{2}$ $\Lambda_c^+$ or $\Lambda_b^0$ baryons, and their $1P$, $2S$, and $2P$ excitations, as well. In this study, we present all the results obtained for all these mentioned excited states in order to be able to make a complete comparison of the masses of each excited state to the observed one.

This work covers the following outline:  Sec.~\ref{Sec2} presents the QCD sum rule calculations for the considered states that are used to get their masses and the current coupling constants. Sec.~\ref{Sec3} provides the details of  numerical analyses to obtain the numerical results. In the last section a summary and conclusion is presented.

\section{QCD sum rule Calculations}~\label{Sec2}

The following two-point correlation function serves as the main ingredient for the calculations of the mass and current coupling constant via the QCD sum rule method:
\begin{equation}
\Pi(q)=i\int d^{4}x e^{iq\cdot x}\langle 0|\mathcal{T}\{J(x)\bar{J}(0)\}|0\rangle . \label{eq:CorrF1}
\end{equation}        
where the $J(x)$ corresponds to the interpolating current of the spin-$\frac{1}{2}$ $\Lambda_Q$ baryon formed from quark fields considering quantum numbers of the considered states. $\mathcal{T}$ is the time ordering operator. For the present work, the interpolating current has the following form: 
\begin{eqnarray}
J=\frac{1}{\sqrt{6}}\epsilon^{abc}\Big[2(u^T_a C d_b)\gamma_5Q_c+2\beta(u^T_aC\gamma_5 d_b)Q_c+(u^T_aCQ_b)\gamma_5d_c+\beta(u^T_aC\gamma_5 Q_b)d_c+(Q^T_aC d_b)\gamma_5 u_c+\beta(Q^T_aC\gamma_5 d_b)u_c\Big].\nonumber\\
\label{Eq:int}
\end{eqnarray}
In the interpolating field, we use the quark field $Q$ for the $c$ or $b$ quark, $C$ is the charge conjugation operator, and $a$, $b$, and $c$ are the color indices of the related quark fields. $\beta$ in the Eq.~(\ref{Eq:int}) is an arbitrary mixing parameter, and its working region shall be established from the analyses of the results. Note that the states under consideration all have the same quark content and quantum numbers, therefore they can be interpolated by the same current.

The calculation of the Eq.~(\ref{eq:CorrF1}) proceeds in two paths. The first path includes using the above interpolating field $J$ explicitly in the Eq.~(\ref{eq:CorrF1}) and getting the results in terms of the  quark-gluon condensates, QCD coupling constant, the masses of the quarks, etc which are the QCD degrees of freedom. The second path requires the calculation of the same correlator, the Eq.~(\ref{eq:CorrF1}), in terms of the hadronic parameters, i.e., masses, current coupling constants, etc. Therefore these two representations are called as the QCD  and the hadronic sides, respectively. Equating the results obtained from each side via a dispersion relation, after isolating the coefficients of the same Lorentz structures on each side, gives a relation to obtain the physical quantities under quest.

For the QCD side, we insert the Eq.~(\ref{Eq:int}) inside the Eq.~(\ref{eq:CorrF1}) and apply possible contractions between proper quark fields using Wick's theorem. The contraction step turns the result into an expression that contains heavy and light quark propagators. These propagators are:
\begin{eqnarray}
S_{q,}{}_{ab}(x)&=&i\delta _{ab}\frac{\slashed x}{2\pi ^{2}x^{4}}-\delta _{ab}%
\frac{m_{q}}{4\pi ^{2}x^{2}}-\delta _{ab}\frac{\langle \overline{q}q\rangle
}{12} +i\delta _{ab}\frac{\slashed xm_{q}\langle \overline{q}q\rangle }{48}%
-\delta _{ab}\frac{x^{2}}{192}\langle \overline{q}g_{\mathrm{s}}\sigma
Gq\rangle +i\delta _{ab}\frac{x^{2}\slashed xm_{q}}{1152}\langle \overline{q}%
g_{\mathrm{s}}\sigma Gq\rangle  \notag \\
&&-i\frac{g_{\mathrm{s}}G_{ab}^{\alpha \beta }}{32\pi ^{2}x^{2}}\left[ %
\slashed x{\sigma _{\alpha \beta }+\sigma _{\alpha \beta }}\slashed x\right]
-i\delta _{ab}\frac{x^{2}\slashed xg_{\mathrm{s}}^{2}\langle \overline{q}%
q\rangle ^{2}}{7776} ,  \label{Eq:qprop}
\end{eqnarray}%
and
\begin{eqnarray}\label{proheavy} 
S_{Q,ab}(x) \!\!\! &=& \!\!\! {\delta _{ab}m_Q^2 \over 4 \pi^2} {K_1(m_Q\sqrt{-x^2}) \over \sqrt{-x^2}} -
i \delta _{ab}{m_Q^2 \rlap/{x} \over 4 \pi^2 x^2} K_2(m_Q\sqrt{-x^2})
-ig_s \int {d^4k \over (2\pi)^4} e^{-ikx} \int_0^1
du \Bigg[ {\rlap/k+m_Q \over 2 (m_Q^2-k^2)^2} G^{\mu\nu}_{ab} (ux)
\sigma_{\mu\nu}\nonumber \\
&& +
{u \over m_Q^2-k^2} x_\mu G^{\mu\nu}_{ab} (ux)\gamma_\nu \Bigg].
\end{eqnarray}
Here, $K_{\nu}$ is the Bessel function of the second kind and  $G_{ab}^{\alpha\beta}=G_A^{\alpha\beta}t^A_{ab}$ is the gluon field strength tensor with $A=1,~2,\cdots,8$ and $t^A=\lambda^A/2$. We use the propagators and make Fourier transformation to convert the result to momentum space and Borel transformation to suppress the contribution of the higher states and continuum. To further suppress these contributions, we use continuum subtraction and finally get the results in terms of QCD degrees of freedom which is a lengthy result. So, we here only focus on their analyses without giving them explicitly. The result of the calculation gives the spectral density $\rho(s)=\frac{1}{\pi}\mathrm{Im}[\Pi^{\mathrm{QCD}}]$ that is used in the following relation
\begin{eqnarray}
\tilde{\Pi}^{\mathrm{QCD}}(s_0,M^2)=\int_{(m_Q+m_u+m_d)^2}^{s_0}dse^{-\frac{s}{M^2}}\rho(s)+\Gamma(M^2),
\label{Eq:Cor:QCD}
\end{eqnarray}
where, as we stated above, we do not present the very lengthy functions $ \rho(s) $ and $ \Gamma(M^2) $ here.

Next, the standard sum rule procedure requires the calculation of Eq.~(\ref{eq:CorrF1}) in terms of hadronic parameters. To fulfill this part,  complete sets of intermediate states for the resonances under study are placed inside the correlation function. As a result of this step, we get
\begin{eqnarray}
\Pi^{\mathrm{Had}}(q)&=&\frac{\langle0|J(0)|\Lambda_Q(q,s)\rangle\langle\Lambda_Q(q,s)|\bar{J}(0)|0\rangle}{m^2-q^2}
+\frac{\langle0|J(0)|\tilde{\Lambda}_Q(q,s)\rangle\langle\tilde{\Lambda}_Q(q,s)|\bar{J}(0)|0\rangle}{\tilde{m}^2-q^2}
+\frac{\langle0|J(0)|\Lambda_Q'(q,s)\rangle\langle\Lambda_Q'(q,s)|\bar{J}(0)|0\rangle}{m'{}^2-q^2}+\nonumber\\
&+&\frac{\langle0|J(0)|\tilde{\Lambda}_Q'(q,s)\rangle\langle\tilde{\Lambda}_Q'(q,s)|\bar{J}(0)|0\rangle}{\tilde{m}'{}^2-q^2}\cdots.
\label{Eq:cor:Phys}
\end{eqnarray}
The $|\Lambda_Q(q,s)\rangle$, $|\tilde{\Lambda}_Q(q,s)\rangle$, $|\Lambda_Q'(q,s)\rangle$, and $|\tilde{\Lambda}_Q'(q,s)\rangle$ are used for the one-particle states of the ground state $(1S)$, and its radial or orbital excitation states, i.e., $1P$, $2S$ and $2P$ states, respectively, with their corresponding masses, $m$, $\tilde{m}$, $m'$, and  $\tilde{m}'$. The $\cdots$ is used for higher states and continuum contributions. The matrix elements in the Eq.~(\ref{Eq:cor:Phys}) are given in terms of the current coupling constants, $\lambda$, $\tilde{\lambda}$, $\lambda'$, and $\tilde{\lambda}'$, and have the following forms:
\begin{eqnarray}
\langle 0|J(0)|\Lambda_Q(q,s)\rangle&=&\lambda u(q,s),\nonumber\\
\langle 0|J(0)|\tilde{\Lambda}_Q(q,s)\rangle&=&\tilde{\lambda}\gamma_5 \tilde{u}(q,s),\nonumber\\
\langle 0|J(0)|\Lambda_Q'(q,s)\rangle&=&\lambda' u'(q,s),\nonumber\\
\langle 0|J(0)|\tilde{\Lambda}_Q'(q,s)\rangle&=&\tilde{\lambda}'\gamma_5 \tilde{u}'(q,s),
\label{Eq:Matrixelm}
\end{eqnarray}  
where $u(q,s)$, $\tilde{u}(q,s)$, $u'(q,s)$, and $\tilde{u}'(q,s)$ are the Dirac spinors of considered states obeying the following summation rule
\begin{eqnarray}
\sum_{s}u(q,s)\bar{u}(q,s)=(\not\!q+m).
\label{Eq:sumspin}
\end{eqnarray}  
The matrix elements in Eq.~(\ref{Eq:Matrixelm}) are placed inside the Eq.~(\ref{Eq:cor:Phys}), and the summations over the spins are applied using Eq.~(\ref{Eq:sumspin}) to get
\begin{eqnarray}
\Pi^{\mathrm{Had}}(q)=\frac{\lambda^2(\not\!q+m)}{m^2-q^2}+\frac{\tilde{\lambda}^2(\not\!q-\tilde{m})}{\tilde{m}^2-q^2}+\frac{\lambda'^2(\not\!q+m')}{m'{}^2-q^2}+\frac{\tilde{\lambda}'^2(\not\!q-\tilde{m}')}{\tilde{m}'^2-q^2}+\cdots.
\label{Eq:cor:Phys1}
\end{eqnarray}
Finally, the Borel transformation is applied to suppress the contributions of higher states and continuum, and the physical side takes its final form as:
\begin{eqnarray}
\tilde{\Pi}^{\mathrm{Had}}(q)=\lambda^2(\not\!q+m)e^{-\frac{m^2}{M^2}}+\tilde{\lambda}^2(\not\!q-\tilde{m})e^{-\frac{\tilde{m}^2}{M^2}}+\lambda'^2(\not\!q+m')e^{-\frac{m'{}^2}{M^2}}+\tilde{\lambda}'^2(\not\!q-\tilde{m}')e^{-\frac{\tilde{m}'^2}{M^2}}\cdots.
\label{Eq:cor:Fin}
\end{eqnarray}
We represent the Borel transformed correlation function by $\tilde{\Pi}^{\mathrm{Had}}(q)$ and contributions of higher states and continuum by the $\cdots$.

The results obtained from the QCD  and hadronic sides are matched to give the relations to obtain the physical quantities under question. On both sides, we have two Lorentz structures, $\not\!q$ and  the unit matrix $I$. To make this match, we isolate the coefficients of the same structures from both sides and equate them to each other, which leads us to
\begin{eqnarray}
\lambda^2 e^{-\frac{m^2}{M^2}}+\tilde{\lambda}^2e^{-\frac{\tilde{m}^2}{M^2}}+\lambda'^2e^{-\frac{m'{}^2}{M^2}}+\tilde{\lambda}'{}^2e^{-\frac{\tilde{m}'{}^2}{M^2}}=\tilde{\Pi}^{\mathrm{QCD}}_{\not\!q}(s_0,M^2),
\label{Eq:cor:match1}
\end{eqnarray}
and
\begin{eqnarray}
\lambda^2 m e^{-\frac{m^2}{M^2}}-\tilde{\lambda}^2\tilde{m}e^{-\frac{\tilde{m}^2}{M^2}}+\lambda'^2m'e^{-\frac{m'{}^2}{M^2}}-\tilde{\lambda}'{}^2\tilde{m}'e^{-\frac{\tilde{m}'{}^2}{M^2}}=\tilde{\Pi}^{\mathrm{QCD}}_{I}(s_0,M^2),
\label{Eq:cor:match2}
\end{eqnarray}
considering the coefficient of the $\not\!q$ and $I$, respectively.

In the analyses, we work with the  Lorentz structure $\not\!q$ and take into account the states one by one in such a way: we first consider the ground state, the first term on the left-hand side of Eq.~(\ref{Eq:cor:match1}), and take the remaining terms inside the continuum, i.e., the ground state+continuum scheme. To obtain the mass of the ground state, we take the derivative of the Eq.~(\ref{Eq:cor:match1}) with respect to $-\frac{1}{M^2}$ and then divide the result by again Eq.~(\ref{Eq:cor:match1}). Hence,
\begin{eqnarray}
m^2=\frac{[\tilde{\Pi}^{\mathrm{QCD}}_{\not\!q}(s_0,M^2)]'}{\tilde{\Pi}^{\mathrm{QCD}}_{\not\!q}(s_0,M^2)},
\label{Eq:mass:Groundstate}
\end{eqnarray}
where $[\tilde{\Pi}^{\mathrm{QCD}}_{\not\!q}(s_0,M^2)]'=\frac{d}{d(-\frac{1}{M^2})}[\tilde{\Pi}^{\mathrm{QCD}}_{\not\!q}(s_0,M^2)]$. The corresponding current coupling constant is obtained directly from Eq.~(\ref{Eq:cor:match1}) using the result of mass prediction as
\begin{eqnarray}
\lambda^2 = e^{\frac{m^2}{M^2}}\tilde{\Pi}^{\mathrm{QCD}}_{\not\!q}(s_0,M^2).
\label{Eq:cor:currentcouplground}
\end{eqnarray}
The spectral parameters of the excited states are calculated following similar steps. For the $1P$ state, the first two terms in Eq.~(\ref{Eq:cor:match1}) are considered  and the remaining two terms are put into the continuum, and  the $\mathrm{ground ~ state}+1P~\mathrm{state}+\mathrm{continuum~scheme}$ is used. In this step,   the results obtained from the ground state for the mass and current coupling constant are substituted as inputs to get the properties of the $ 1P $ state. For the $2S$ state, the last term  in Eq.~(\ref{Eq:cor:match1}) is considered as the part of the continuum, that is now we take $\mathrm{ground ~state}+1P~\mathrm{state}+2S~\mathrm{state}+\mathrm{continuum ~scheme}$, and the results obtained for the ground state and first excited state are taken as inputs. And finally, $\mathrm{ground state}+1P~\mathrm{state} +2S~\mathrm{state}+2P~\mathrm{state}+\mathrm{continuum~scheme}$ is considered similarly to get the results for the $2P$ excited state. The numerical outcomes of all these steps are presented in the next section.

\section{Numerical Analyses}\label{Sec3}

After the calculation of the QCD sum rule expressions giving the masses and current coupling constants, these expressions are used to obtain their numerical values with the input parameters given in Table~\ref{tab:Param}.
\begin{table}[tbp]
\begin{tabular}{|c|c|}
\hline\hline
Parameters & Values \\ \hline\hline
$m_{c}$                                     & $1.27\pm 0.02~\mathrm{GeV}$ \cite{ParticleDataGroup:2020ssz}\\
$m_{b}$                                     & $4.18^{+0.03}_{-0.02}~\mathrm{GeV}$ \cite{ParticleDataGroup:2020ssz}\\
$m_{u}$                                     & $2.16^{+0.49}_{-0.26}~\mathrm{MeV}$ \cite{ParticleDataGroup:2020ssz}\\
$m_{d}$                                     & $4.67^{+0.48}_{-0.17}~\mathrm{MeV}$ \cite{ParticleDataGroup:2020ssz}\\
$\langle \bar{q}q \rangle (1\mbox{GeV})$    & $(-0.24\pm 0.01)^3$ $\mathrm{GeV}^3$ \cite{Belyaev:1982sa}  \\
$m_{0}^2 $                                  & $(0.8\pm0.1)$ $\mathrm{GeV}^2$ \cite{Belyaev:1982sa}\\
$\langle \frac{\alpha_s}{\pi} G^2 \rangle $ & $(0.012\pm0.004)$ $~\mathrm{GeV}^4 $\cite{Belyaev:1982cd}\\
$\langle g_s^3 G^3 \rangle $                & $ (0.57\pm0.29)$ $~\mathrm{GeV}^6 $\cite{Narison:2015nxh}\\
\hline\hline
\end{tabular}%
\caption{The parameters used as input in the analyses.}
\label{tab:Param}
\end{table}
The main goal of the present work is the investigation of the possible quantum numbers for the newly observed $\Lambda_c(2910)^+$ state, which was reported as possible $2P$ excitation by the Belle Collaboration~\cite{Belle:2022hnm}. In addition, we also consider its bottom counterpart to get a mass prediction for a possible spin-$\frac{1}{2}$ $2P$ excitation of the $\Lambda_b^0$ state. We consider the excited states $1P$, $2S$, and $2P$ of the $\Lambda_c^+$ baryon to be able to compare them all with the observed states to fix their quantum numbers. The predictions for the ground, $1P$, and $2S$ states of both the spin-$\frac{1}{2}$ $\Lambda_c^+$ and $\Lambda_b^0$ baryons  were all analyzed in our previous work,  Ref.~\citep{Azizi:2020ljx}, and the predicted results are given in the table of results, Table~\ref{tab:results}, and we address the analyses here briefly for completeness of the discussion of the new prediction. Besides the input parameters,  three more auxiliary parameters enter the calculations which are the parameter $\beta$, the threshold parameter $s_0$, and the Borel parameter $M^2$. For the calculations, we need their proper working regions, and these working regions are determined by the criteria that are the standard application of the method used. These criteria include the dominance of the contributions coming from the interested states compared to the higher states and continuum, the moderate variance of the results as functions of these auxiliary parameters, and convergence of the operator product expansion (OPE).

\begin{figure}[h!]
\begin{center}
\includegraphics[totalheight=5cm,width=7cm]{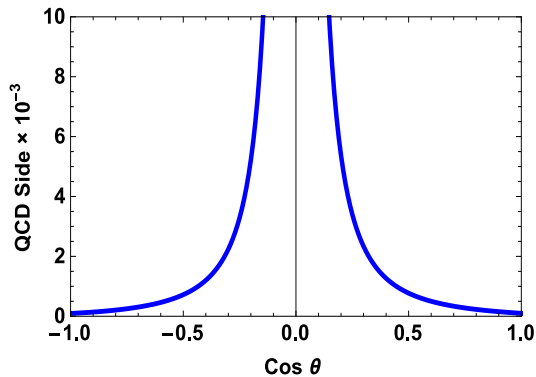}
\end{center}
\caption{The variation of the QCD side (in units of $\mathrm{GeV}^{6}$) for the $2P$ excitation of $\Lambda_c$ baryon as a function of $\cos\theta$.}
\label{gr:costheta}
\end{figure}

First of all, we would like to determine the working  interval for the mixing parameter $\beta$.  This parameter can take values from $- \infty $ to $+\infty$.  For simplicity, we define  $\beta=\tan\theta$, and  discuss the variations of the results with respect to  $\cos\theta$ in the interval $[-1,1]$ to span the whole region. We plot the  QCD side of the calculations in terms of $\cos\theta$ and look for the intervals that the changes with respect to this parameter is minimal.  Figure~\ref{gr:costheta} shows such a parametric plot for the QCD side of the $2P$ excitation of the $\Lambda_c^+$ state at average values of the Borel parameter and continuum threshold to be discussed later.  From this figure and analyses of the results,  we find that, in the intervals   
\begin{eqnarray}
-1.0 < \cos\theta < -0.5~~~ \mathrm{and}~~~0.5 < \cos\theta < 1.0,
\label{costheta}
\end{eqnarray}
the variations of the results are relatively small. We consider these intervals as working  windows of  $\cos\theta$  for both the $\Lambda_c^+$ and $\Lambda_b^0$  channels. The residual dependencies of the results  on this parameter appear as uncertainties in the values of the physical observables under consideration. 

\begin{figure}[h!]
\begin{center}
\includegraphics[totalheight=5cm,width=7cm]{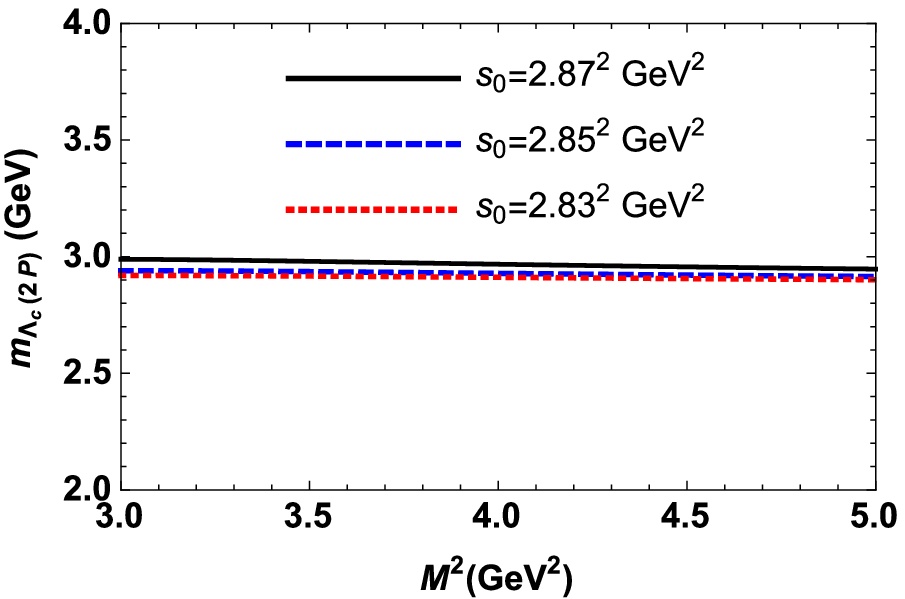}
\includegraphics[totalheight=5cm,width=7cm]{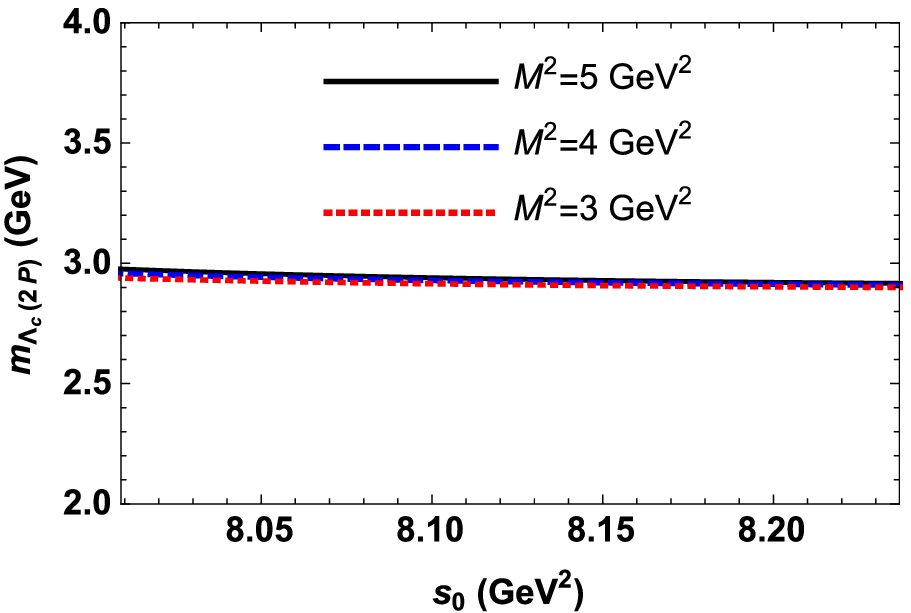}
\end{center}
\caption{\textbf{Left:} The dependence of the mass $\widetilde{m}'$ of the $2P$ excitation of $\Lambda_c$ baryon on Borel parameter $M^2$.
\textbf{Right:} The dependence of the mass $\widetilde{m}'$ of the $2P$ excitation of $\Lambda_c$ baryon on threshold parameter $s_0$.}
\label{gr:mLamdac}
\end{figure}

\begin{figure}[h!]
\begin{center}
\includegraphics[totalheight=5cm,width=7cm]{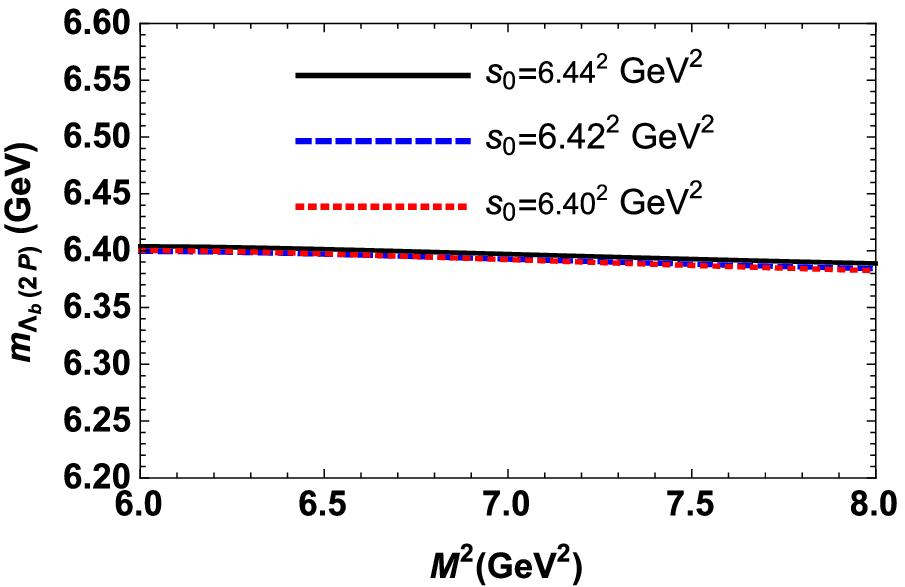}
\includegraphics[totalheight=5cm,width=7cm]{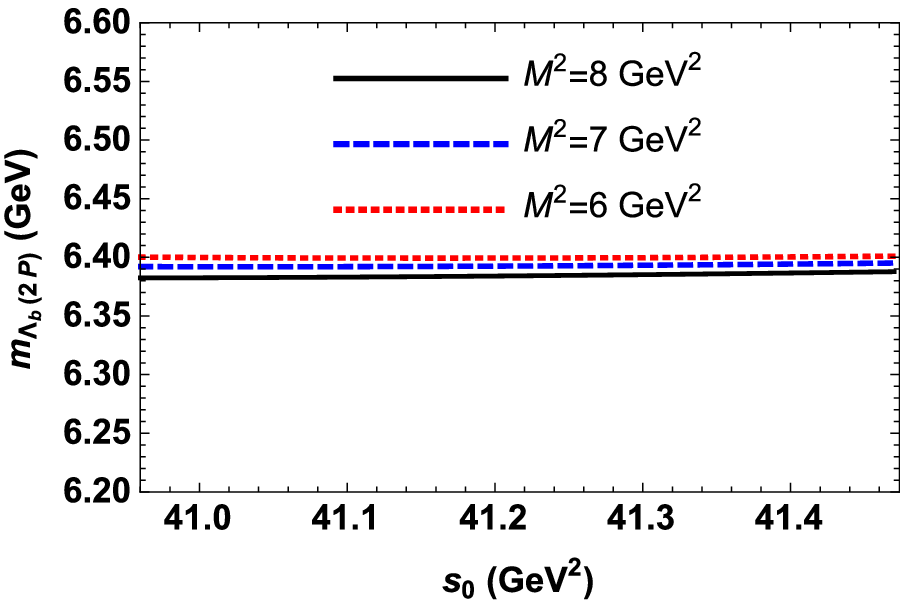}
\end{center}
\caption{\textbf{Left:} The dependence of the mass $\widetilde{m}'$ of the $2P$ excitation of $\Lambda_b$ baryon on Borel parameter $M^2$.
\textbf{Right:} The dependence of the mass $\widetilde{m}'$ of the $2P$ excitation of $\Lambda_b$ baryon on threshold parameter $s_0$.}
\label{gr:mLamdab}
\end{figure}

For the working regions of the Borel parameter, the pole dominance over the continuum states and convergence of the OPE results are imposed and the regions satisfying these requirements and giving mild dependence on these auxiliary parameters are established.  As for the threshold parameter, due to the notion that this parameter has a relation to the energy of the next excited state, its intervals are different for the calculation of each state. For instance, in the case of the calculation of the ground state's mass and current coupling constant, its interval is fixed as $s_0~\epsilon~[2.53^2,2.57^2]$ for $\Lambda_c^+$ ground state considering the ground state+continuum scheme mentioned in the previous section. Afterward, the obtained results are used as input in the calculation of the $1P$ state's mass and current coupling constant with a new proper choice of threshold parameter considering the ground state+$1P$~state+continuum scheme. And these steps are repeated consecutively for the $2S$ and for the $2P$ states of $\Lambda_c^+$  until their corresponding quantities are attained. Table~\ref{tab:results} presents these parameters, $M^2$ and $s_0$, and the results of masses and current coupling constants for all considered states. In Figures~\ref{gr:mLamdac} and \ref{gr:mLamdab} the dependence of the results obtained for the masses of the $2P$ excitations of the $\Lambda_c^+$ and the $\Lambda_b^0$ states as functions of the Borel parameter $M^2$ and $s_0$ at the average value of the parameter $\beta$ are given. These figures depict stability of the results, as expected by the criteria of the QCD sum rule method, with the variations of these parameters in their working intervals. 
\begin{table}[]
\begin{tabular}{|c|c|c|c|c|c|}
\hline
   Particle  & State &$M^2~(\mathrm{GeV^2})$&$s_0~(\mathrm{GeV^2})$  & Mass~(MeV) & $\lambda~(\mathrm{GeV^3})$ \\ \hline\hline

\multirow{3}{*}{} &$\Lambda_c(\frac{1}{2}^+)(1S)$ &$3.0-5.0$& $2.53^2-2.57^2$& $2282.42\pm28.38$ & $0.022\pm0.001$ \\ \cline{2-6} 
  $\Lambda_c^+$   &$\Lambda_c(\frac{1}{2}^-)(1P)$ &$3.0-5.0$& $2.63^2-2.67^2$& $2592.36\pm104.85$ & $0.014\pm0.004$ \\ \cline{2-6} 
                  &$\Lambda_c(\frac{1}{2}^+)(2S)$ &$3.0-5.0$& $2.73^2-2.77^2$& $2765.52\pm77.60$ & $0.016\pm0.005$ \\
\cline{2-6} 
                  &$\Lambda_c(\frac{1}{2}^-)(2P)$ &$3.0-5.0$& $2.83^2-2.87^2$& $2935.23\pm134.70$ & $0.018\pm0.003$ \\
                  \hline\hline       
\multirow{3}{*}{} &$\Lambda_b(\frac{1}{2}^+)(1S)$ &$6.0-8.0$& $5.86^2-5.90^2$& $5611.47\pm27.47$ & $0.042\pm0.003$  \\ \cline{2-6} 
   $\Lambda_b^0$  &$\Lambda_b(\frac{1}{2}^-)(1P)$ &$6.0-8.0$& $5.92^2-5.96^2$& $5910.56\pm84.54$ & $0.020\pm0.008$  \\ \cline{2-6} 
                  &$\Lambda_b(\frac{1}{2}^+)(2S)$ &$6.0-8.0$& $6.18^2-6.22^2$& $6073.65\pm93.22$ & $0.051\pm0.007$ \\             
\cline{2-6} 
                  &$\Lambda_b(\frac{1}{2}^-)(2P)$ &$6.0-8.0$& $6.40^2-6.44^2$& $6392.75\pm81.00$ & $0.053\pm0.005$ \\             
                  \hline
\end{tabular}
\caption{The predictions for the masses and  the current coupling constants and the auxiliary parameters used in their analyses.}
\label{tab:results}
\end{table}
The errors present in the results come mainly from the uncertainties of the auxiliary parameters. The uncertainties of other input parameters also contribute to these errors.  It should also be noted that, among the states under study, the $ \Lambda_c(\frac{1}{2}^-)(1P)$ lies very close to the $ \Sigma_c(2455)\pi $ threshold. In principle $ \Sigma_c(2455)\pi $ state also contributes and modifies the hadronic side of the sum rule obtained for  the  $\Lambda_c$ resonances in the present study. However, it is reasonable to neglect such contributions for narrow resonances.  Our estimation for the order of  uncertainties imposing from such an approximation  are  respectively about $ 2\% $ and  $ 7\% $ for the mass and current coupling constant, which were added to the corresponding numerical results in Table~\ref{tab:results}.

\section{Conclusion}

Recently the Belle Collaboration announced a new exited charm baryon state which was reported to be a possible candidate for $\Lambda_c(\frac{1}{2}^-,2P)$ state and decaying into $\Sigma_c(2455)^{0,++}\pi^{\pm}$~\cite{Belle:2022hnm}. Mass and width for the particle were given as $2913.8\pm 5.6 \pm 3.8~\mathrm{MeV}/c^2$ and $(51.8\pm 20.0 \pm 18.8)~\mathrm{MeV}$, respectively. This observation motivated us to analyze the spin-$\frac{1}{2}$ $\Lambda_c^+$ state for its excitations to search whether the observed state is possibly a $2P$ excitation of the ground state $\Lambda_c^+$. To predict the possible quantum numbers of this state, we performed a QCD sum rule analysis for the mass of spin-$\frac{1}{2}$ $\Lambda_c^+$ states and take into account not only its $2P$ excitation but also $1P$ and $2S$ excitations to have a complete comparison. In addition, we provided a prediction for a $2P$ spin-$\frac{1}{2}$ $\Lambda_b^0$ state to supply insight for its possible future observation.

For the discussion of the results for ground $(1S)$, $1P$, and $2S$ states we refer the readers to our previous work~\cite{Azizi:2020ljx}, where  the predicted masses for the $\Lambda_c(\frac{1}{2})$ $1S$, $1P$, and $2S$ states were   $m(1S)=2282.42\pm28.38$~MeV, $m(1P)=2592.36\pm 53.01$~MeV and $m(2S)=2765.52\pm 22.29$~MeV, respectively. These results are in agreement, within the errors, with the other theoretical predictions~\cite{Capstick:1986bm,Ebert:2007nw,Ebert:2011kk,Migura:2006ep,Roberts:2007ni,Chen:2016iyi,Chen:2014nyo,Shah:2016nxi,Valcarce:2008dr,Lu:2016ctt,Yang:2017qan}. The main focus of this work is the determination of the quantum numbers of the newly observed state reported by the Belle Collaboration~\cite{Belle:2022hnm}. Comparison of the  reported mass of this state with our findings for the ground, $1P$, $2S$, and $2P$ states of spin-$\frac{1}{2}$ $\Lambda_c^+$ state indicates that $1P$ and $2S$ states' masses are lower than that of the observed one. On the other hand, the mass of the $2P$ state obtained in the present study,  $m=2935.23\pm134.70$~MeV, has good consistency with the observed mass considering its error.  This suggests  the interpretation of the observed state as $2P$ excitation of the $\Lambda_c^+$ state with quantum numbers $J^P=\frac{1}{2}^-$. This result is also in accord with the predictions of the other theoretical works given for $2P$ excited state of $\Lambda_c^+(\frac{1}{2})$ state such as $m=2983$~MeV~\cite{Ebert:2011kk}, $m=2980$~MeV~\cite{Chen:2016iyi}, $m=2989$~MeV~\cite{Chen:2014nyo}, $m=2890$~MeV~\cite{Yoshida:2015tia}, $m=2880$~MeV~\cite{Valcarce:2008dr}, $m=2764-2956$~MeV~\cite{Yang:2017qan} and $m=2988$~MeV~\cite{Yu:2022ymb}. Considering the error range, our prediction is also consistent with those of Refs.~\cite{Ebert:2007nw,Shah:2016nxi,Lu:2016ctt}, which are $m=3062$~MeV,  $m=3017$~MeV and $m=3030$~MeV, respectively as well as $m=2853$~MeV~\cite{Migura:2006ep}, and $m=2816$~MeV~\cite{Roberts:2007ni}. It is, however, larger than $m=2780$~MeV obtained in \cite{Capstick:1986bm}.

For the bottom counterpart, $\Lambda_b^0$, with spin-$\frac{1}{2}$ ground, $1P$, $2S$ states, our  predictions for their masses were $m(1S)=5611.47\pm27.47$~MeV, $m(1P)=5910.56\pm 84.54$~MeV and $m(2S)=6073.65\pm 93.22$~MeV~\cite{Azizi:2020ljx}. These results are consistent with Refs.~\cite{Capstick:1986bm,Ebert:2007nw,Ebert:2011kk,Roberts:2007ni,Chen:2014nyo,Yoshida:2015tia,Valcarce:2008dr} within their errors. For the corresponding $2P$ state, in this work we got the mass as $6392.75\pm81.00$~MeV. Considering its error, this prediction is consistent with the prediction of Ref.~\cite{Ebert:2007nw} as $m=6328$~MeV, Ref.~\cite{Ebert:2011kk} as $m=6326$~MeV and it is larger than the predictions of Ref.~\cite{Capstick:1986bm} as $m=6100$~MeV, Ref.~\cite{Roberts:2007ni} as $m=6180$~MeV, Ref.~\cite{Yoshida:2015tia} as $m=6236$~MeV, Ref.~\cite{Valcarce:2008dr} as $m=6245$~MeV, Ref.~\cite{Yang:2017qan} as $m=6120-6298$~MeV, and Ref.~\cite{Yu:2022ymb} as $m=6238$~MeV. The mass prediction given for this state may shed light on the future experiments in the observation and determination of the quantum numbers of such a possible state.

The results of the analyses of the present work and its comparison with the Belle Collaboration's observation indicates that the newly seen state $\Lambda_c(2910)^+$ can be  considered as the  $2P$ excitation of the spin-$\frac{1}{2}$ $\Lambda_c^+$ state. The present work also provides a mass prediction for its bottom counterpart, namely $2P$ excitation of the spin-$\frac{1}{2}$ $\Lambda_b^0$ state, which is not observed yet and it may be in agenda of different experiments in near future. The values obtained for the residues or current coupling constants can be used as inputs to analyze different interactions and decays of the considered states. Comparisons of the results attained in the present work with the future experimental and other theoretical investigations of the masses, which may be supported also by the decay properties of these states, may help for the exact determination of the quantum numbers of the considered states.




\label{sec:Num}

\end{document}